\documentclass{article}

\usepackage{natbib}

\usepackage[utf8]{inputenc} 
\usepackage[T1]{fontenc}    
\usepackage{hyperref}       
\usepackage{url}            
\usepackage{booktabs}       
\usepackage{amsfonts}       
\usepackage{nicefrac}       
\usepackage{microtype}      
\usepackage{lipsum}
\usepackage{graphicx}
\graphicspath{ {./images/} }
\usepackage{latexsym}
\usepackage{caption}
\usepackage{subcaption}
\usepackage{amssymb}
\usepackage{lipsum}
\usepackage{array}
\usepackage{longtable}
\usepackage{soul}
\usepackage{comment}
\usepackage{amsmath}
\usepackage{svg}
\usepackage{algorithm}
\usepackage{algpseudocode}

\title{A Multi-Agent Systems Approach for Peer-to-Peer Energy Trading in Dairy Farming}

\author{
 Mian Ibad Ali Shah \\
  School of Computer Science\\
  University of Galway, Ireland\\
  \texttt{m.shah7@universityofgalway.ie} \\
   \And
 Abdul Wahid \\
  School of Computer Science\\
  University of Galway, Ireland\\
  \texttt{abdul.wahid@universityofgalway.ie} \\
  \And
 Enda Barrett \\
  School of Computer Science\\
  University of Galway, Ireland\\
  \texttt{enda.barrett@universityofgalway.ie} \\
   \And
 Karl Mason \\
  School of Computer Science\\
  University of Galway, Ireland\\
  \texttt{karl.mason@universityofgalway.ie} \\
}

\begin{document}

{
    \renewcommand{\thefootnote}{\fnsymbol{footnote}}
    \footnotetext{\textit{Proc. of the Artificial Intelligence for Sustainability, ECAI 2023, Eunika et al. (eds.), Sep 30- Oct 1, 2023, https://sites.google.com/view/ai4s. 2023.}}
}

\maketitle

\begin{abstract}
To achieve desired carbon emission reductions, integrating renewable generation and accelerating the adoption of peer-to-peer energy trading is crucial. This is especially important for energy-intensive farming, like dairy farming. However, integrating renewables and peer-to-peer trading presents challenges. To address this, we propose the Multi-Agent Peer-to-Peer Dairy Farm Energy Simulator (MAPDES), enabling dairy farms to participate in peer-to-peer markets. Our strategy reduces electricity costs and peak demand by approximately 30\% and 24\% respectively, while increasing energy sales by 37\% compared to the baseline scenario without P2P trading. This demonstrates the effectiveness of our approach.
\end{abstract}


\section{Introduction}
\label{introduction}

According to Shine et al. \cite{shine2020energy}, global dairy consumption per capita is projected to increase by 19\% by 2050. However, milk production requires significant energy, raising concerns about carbon emissions. To ensure the future sustainability of the dairy industry, energy resources must be used sustainably \cite{ben2017renewable}. An AI system can help reduce emissions and peak demand for electricity in dairy farms.

Multi-agent systems (MAS) have shown promising results in addressing microgrid challenges \cite{elena2022multi}. Performance evaluation models have optimized profit in energy-sharing regions (ESR) \cite{zhou2017performance}. Peer-to-peer (P2P) energy trading involves sharing energy within a microgrid before trading with the retailer \cite{zhou2018evaluation}.

Three types of P2P energy trading exist: centralized, decentralized, and distributed \cite{zhou2020state}. Distributed markets combine aspects of centralized and decentralized markets, employing auction-based mechanisms such as the Double Auction (DA) \cite{zhou2018evaluation}. This research utilizes MAS, P2P energy trading, price advisor, and auction to contribute:

\begin{enumerate}
    \item Development of a P2P energy trading model using MAS to optimize the utilization of renewable energy (RE) sources and minimize reliance of dairy farms on the utility grid to achieve energy sustainability and reduce carbon emissions
    \item Integrating an internal pricing advisor with the auctioneer which makes the decision-making processes more transparent
\end{enumerate}

\section{Related Work}
\label{Literature}
MAS has been widely studied for P2P energy trading due to its potential for financial benefits, scalability, reliability, data security, user satisfaction, peak demand management, and load management \cite{ye2021scalable}. Researchers have explored different approaches to achieve these objectives. Various techniques have been proposed, such as non-cooperative games for dynamic pricing \cite{zhang2019p2p}, reinforcement learning (RL) models \cite{ali2020synergychain}, and deep RL \cite{chen2019realistic} have also been employed.

In our research, we primarily focus on the distributed approach, which offers scalability and autonomy to customers \cite{khorasany2018market}. Various techniques, including DA-MADDPG \cite{qiu2021multi}, aggregated control models \cite{long2018peer}, and MARL models \cite{pu2022peer}, have been proposed for privacy, profit maximization, and market participation. 
Our research consolidates these findings and integrates critical factors such as financial benefits, data security, scalability, user satisfaction, load management, peak demand management, and transparent auction mechanisms into a comprehensive simulation.

\section{Methodology}
\label{methodology}

\begin{figure*}[h!]
    \centering
    \includegraphics[width=1\textwidth]{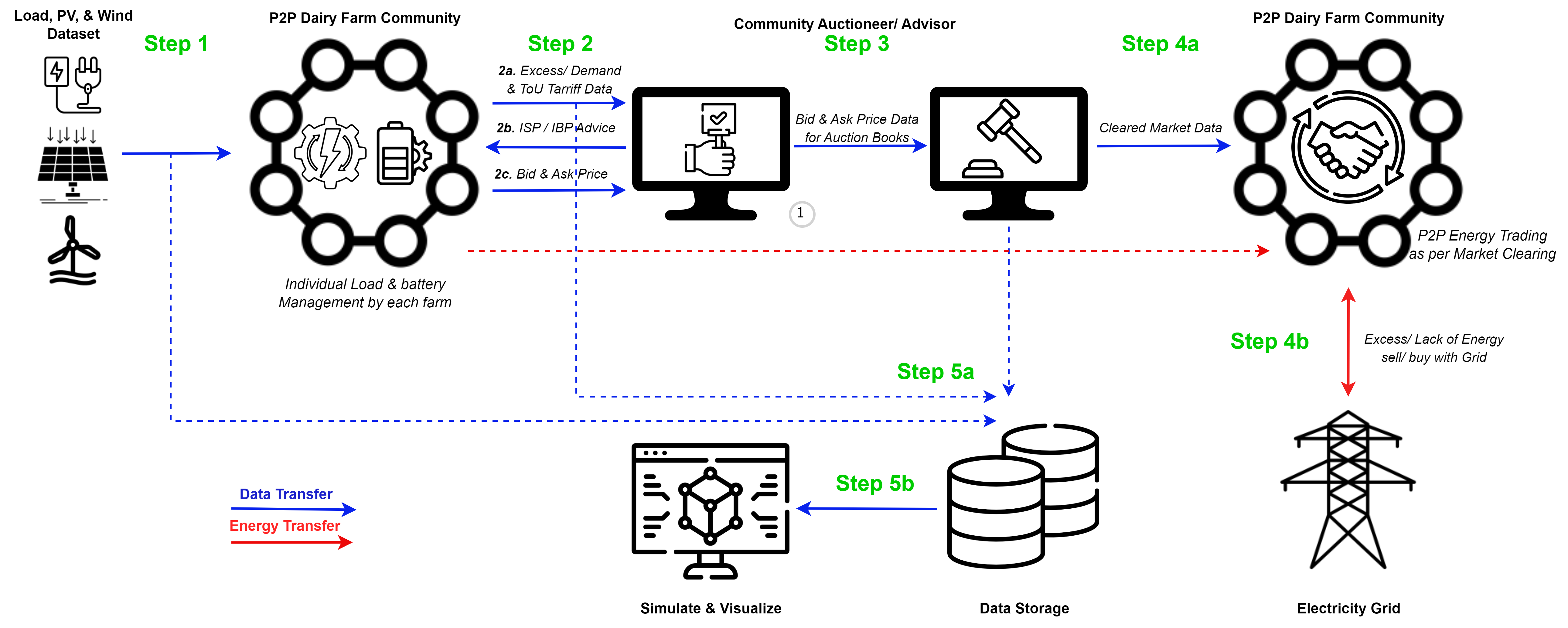}
    \caption{Simulation development steps for distributed P2P energy trading using MAS}
    \label{methodology figure}
\end{figure*}

This study aims to introduce a MAS-based algorithm that facilitates distributed P2P energy trading among dairy farms. The proposed approach leverages MAS to enable dairy farms to sell excess energy produced from renewable sources to other farms in the vicinity. This approach aims to decrease their reliance on the utility grid and promote energy self-reliance. Figure \ref{methodology figure} illustrates the process flow of the simulation model, which is explained in detail in the subsequent sections covering each step.

\subsection{Datasets and Infrastructure}
\label{dataset&infra}

This study combines three datasets: Farm load data from Uski et al. \cite{uski2018microgrid}, PV and wind energy generation data from the System Advisor Model (SAM). The infrastructure supports a scalable number of farms, with diverse sizes and PV/wind systems. Participants are prosumers connected to the utility grid, facilitated by a central auctioneer providing Internal Selling Price (ISP) and Internal Buying Price (IBP) using the Supply Demand Ratio. The infrastructure operates on a distributed peer-to-peer model, with farms sharing generation and load data for centralized market clearance. Various agents, including farms, batteries, and traders, coordinate to meet energy demands effectively.

\subsection{Model Design}
\label{techniques}

The simulator consists of three steps: individual load and battery management, calculation of community's ISP and IBP, and market clearing through auction and energy trading. It can be tailored to simulate any duration, enabling a comprehensive analysis of grid load and potential profit/ loss.

The study proposes a Python-based simulation model for energy generation, consumption, and storage in a farm using renewable sources. It includes tailored load and battery management rules for dairy farming. The model enables energy trading through a P2P network or the grid. Input parameters and variables are used to calculate energy generation, consumption, and storage, considering different scenarios.

\begin{table}[h!]
    \centering
    \label{eq-table}
    \begin{tabular}{@{}l@{}l@{}}
          \\
        1 & $\lambda_{\text{buy}} = \$, \$\$, \$\$\$ = \text{night, day, peak}$ \\
        2 & $B_{uc} = \begin{cases}
        B_{cc}, & \text{if } B_{cc} < \max(B_{dp}) \\
        \max(B_{dp}), & \text{if } B_{cc} \geq \max(B_{dp})
        \end{cases}$ \\
        3 & $E_{\text{tot}} = \begin{cases}
        E_{pv} + E_w + B_{uc}, & \text{$PV = 1$ or $wind = 1$ or $bat = 1$} \\
        0, & \text{otherwise}
        \end{cases}$ \\
        4 & \begin{tabular}[t]{@{}l@{}}$\text{if } E_{\text{tot}} > E_l \text{ \& SoC} < 90:$  $\quad \begin{cases}
        E_{\text{tot}} - E_l < B_{cp}, & \text{charge}=1 \\
        E_{\text{tot}} - E_l > B_{cp}, & \text{charge}=1 \text{ \& sell}=1
        \end{cases}$\end{tabular} \\
        5 & \begin{tabular}[t]{@{}l@{}}$\text{if } E_{\text{tot}} < E_l:$  $\quad \begin{cases}
        SoC > 20 \text{ \& } \lambda_{\text{buy}} = \$, & \text{Buy}=1 \text{ \& charge}=0 \\
        \text{(}SoC<50 \text{ \& } \lambda_{\text{buy}} = \$\text{) or} & \\
        \text{(}SoC<20 \text{ \& } \lambda_{\text{buy}}<\$\$\$, & \text{Buy}=1 \text{ \& charge}=1 \\
        \end{cases}$\end{tabular} \\
        6 & $\text{if } RE = 1 \text{ \& bat}=0: \begin{cases}
        \text{if } E_{\text{tot}} > E_l, & \text{sell}=1 \\
        \text{if } E_{\text{tot}} < E_l, & \text{buy}=1 \\
        \end{cases}$ \\
        7 & \begin{tabular}[t]{@{}l@{}}$\text{if } RE=0 \text{ \& } \text{bat}=1:$  $\quad \begin{cases} 
        \text{if } SoC > 20 \text{ and } \lambda_{\text{buy}} = \$ \text{, Buy}=1 \text{ and charge}=0\\
        \text{if(} SoC < 20 \text{ and } \lambda_{\text{buy}} < \$\$\$ \text{) , Buy}=1 \text{ and charge}=1
        \end{cases}$\end{tabular} \\
        8 & $SoC = \begin{cases}
        SoC + \frac{B_{\text{ccap}}}{B_c} \times 100, & \text{charge}=1 \\
        SoC - B_{\text{dp}}, & \text{discharge}=1
        \end{cases}$ \\
        9 & $B_{\text{ccap}} = \begin{cases}
        \max(B_{\text{ccap}}), & \text{(} RE = 1 \text{ \& } E_{\text{tot}} - E_l > \max(B_{\text{ccap}}) \text{) or } RE = 0 \\
        E_{\text{tot}} - E_l, & RE = 1 \text{ \& } E_{\text{tot}} - E_l < \max(B_{\text{ccap}})
        \end{cases}$ \\
        10 & $B_{\text{dp}} = \frac{B_{uc}}{B_c}\times 100$ \\
        11 & $E_s = \begin{cases}
        E_{\text{tot}} - E_l, & \text{SoC} \geq 90 \\
        (E_{\text{tot}} - E_l) - B_{\text{ccap}}, & (E_{\text{tot}} - E_l) > B_{\text{ccap}} \text{ \& SoC} < 90
        \end{cases}$ \\
        12 & $E_b = \begin{cases}
        E_l, & \text{RE} = 0 \text{ \& bat} = 0 \\
        E_l - (E_{\text{tot}} + B_{uc}), & E_{\text{tot}} < E_l \text{ \& } \lambda_{\text{buy}} > \$ \text{ \& SoC} > 20 \\
        (E_l - E_{\text{tot}}) + B_{\text{ccap}}, & \lambda_{\text{buy}} < \$\$\text{ \& SoC} \leq 50 \\
        E_l - E_{\text{tot}}, & \text{RE} = 1 \text{ \& bat} = 0 \\
        E_l - B_{uc}, & \text{RE} = 0 \text{ \& bat} = 1 \text{ \& SoC} > 20 \text{ \& } \lambda_{\text{buy}} > \$ \\
        E_l + B_{\text{ccap}}, & \text{RE} = 0 \text{ \& bat}=1 \text{ \& SoC}\leq20 \text{ \& } \lambda_{\text{buy}} < \$\$\$
        \end{cases}$ \\
        13 & $SDR = \frac{TSP}{TBP}$ \\
        14 & $\text{ISP} = \begin{cases}
        \frac{\lambda_{\text{sell}} \times \lambda_{\text{buy}}}{(\lambda_{\text{buy}} - \lambda_{\text{sell}}) \times \text{SDR} + \lambda_{\text{sell}}}, & 0 \leq \text{SDR} \leq 1 \\
        \lambda_{\text{sell}}, & \text{SDR} > 1
        \end{cases}$ \\
        15 & $\text{IBP} = \begin{cases}
        \text{ISP} \times \text{SDR} + \lambda_{\text{buy}} \times \text{(1-SDR)}, & 0 \leq \text{SDR} \leq 1 \\
        \lambda_{\text{buy}}, & \text{SDR} > 1
        \end{cases}$ \\
    \end{tabular}
\end{table}

Equation 1 defines the energy tariff purchased from the grid, denoted as $\lambda_{buy}$, with three levels: \$, \$\$, and \$\$\$, representing night, day, and peak hours, respectively. Equation 2 calculates the present usable battery capacity ($B_{uc}$) based on the current battery capacity ($B_{cc}$) and the maximum battery discharge capacity ($\max(B_{dp})$). If $B_{cc}$ is less than $\max(B_{dp})$, $B_{uc}$ is equal to $B_{cc}$; otherwise, it is set to $\max(B_{dp})$. Equation 3 evaluates the total energy generation of the farm ($E_{tot}$) by summing up the energy generated by the PV system ($E_{pv}$), wind turbine ($E_w$), and $B_{uc}$. If any of these sources do not generate energy, $E_{tot}$ is zero.

Equation 4 determines the optimal battery operation when renewable energy resources and batteries are available on the farm. If $E_{tot}$ is greater than $E_l$ (farm load) and the current battery percentage ($SoC$) is less than 90\%, the battery is charged or discharged based on the difference between $E_{tot}$ and $E_l$, and the charging capacity ($B_{ccap}$). If the excess energy is insufficient to charge the battery, only the battery is charged; otherwise, the remaining energy is sold in the market.

Equation 5 determines whether to charge or discharge the battery and whether to purchase energy from the grid based on $E_{tot}$, $SoC$, and the energy tariff ($\lambda_{buy}$). The conditions in this equation guide the decision-making process. Equation 6 decides whether to buy or sell energy based on $E_{tot}$ and $E_l$, considering the presence or absence of a battery. Equation 7 determines whether energy should be purchased externally and whether the battery should be charged, based on $SoC$ and $\lambda_{buy}$.

Equation 8 updates the battery percentage ($SoC$) based on its charging or discharging status. Equations 9 and 10 calculate the battery charging capacity ($B_{ccap}$) and discharge percentage ($B_{dp}$) using the given conditions. Equations 11 and 12 determine the excess energy available ($E_s$) and the amount of energy bought ($E_b$), respectively, based on various scenarios involving renewable energy generation, battery capacity and percentage, energy tariffs, and load demand.

After implementing individual load and battery management at farms, the subsequent action is to engage in transactions involving the sale or purchase of surplus or deficient energy to or from the community or grid, respectively.

IBP and ISP are computed using the SDR technique \cite{liu2017energy} for energy trading. Internal prices consider the feed-in tariff, utility grid electricity prices, and economic balance. As seen in Equations 14 and 15, ISP and IBP are determined through a piecewise function based on the SDR. The pricing strategy depends on the SDR, guiding the selling price between $\lambda_{sell}$ and $\lambda_{buy}$ (feed-in tariff and time of use) for a balanced supply-demand ratio. Prosumers aim to increase the SDR by reducing consumption when it is small, while sellers and buyers adjust consumption accordingly when the SDR is large.

\begin{figure}[h!]
    \centering
    \begin{subfigure}{0.49\textwidth}
        \includegraphics[width=\linewidth]{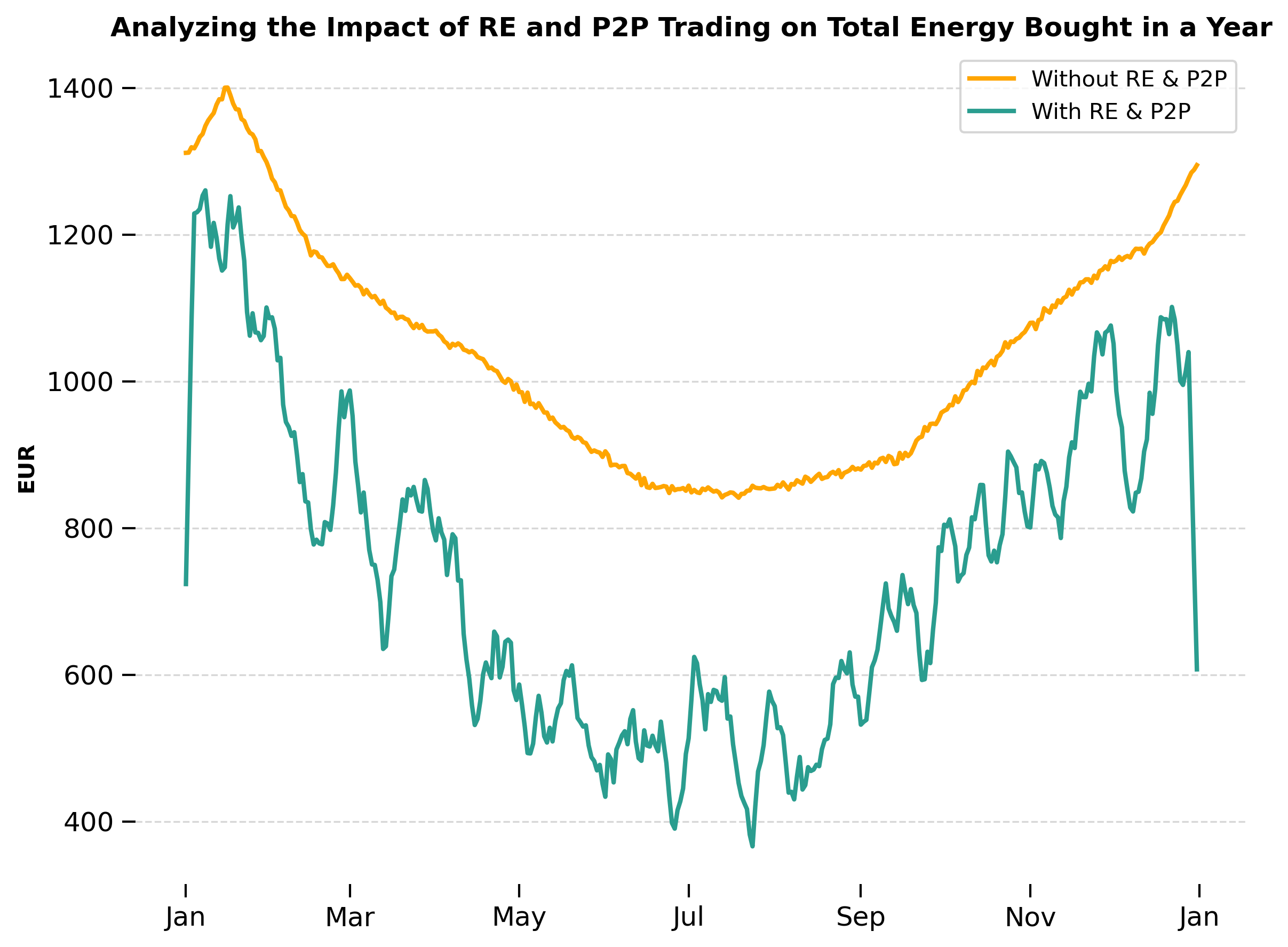}
        \caption{Comparison of Energy Purchased by Farm Community: P2P and RE vs Non-P2P and Non-RE Sources}
        \label{boughtwoRE}
    \end{subfigure}
    \hfill
    \begin{subfigure}{0.49\textwidth}
        \includegraphics[width=\linewidth]{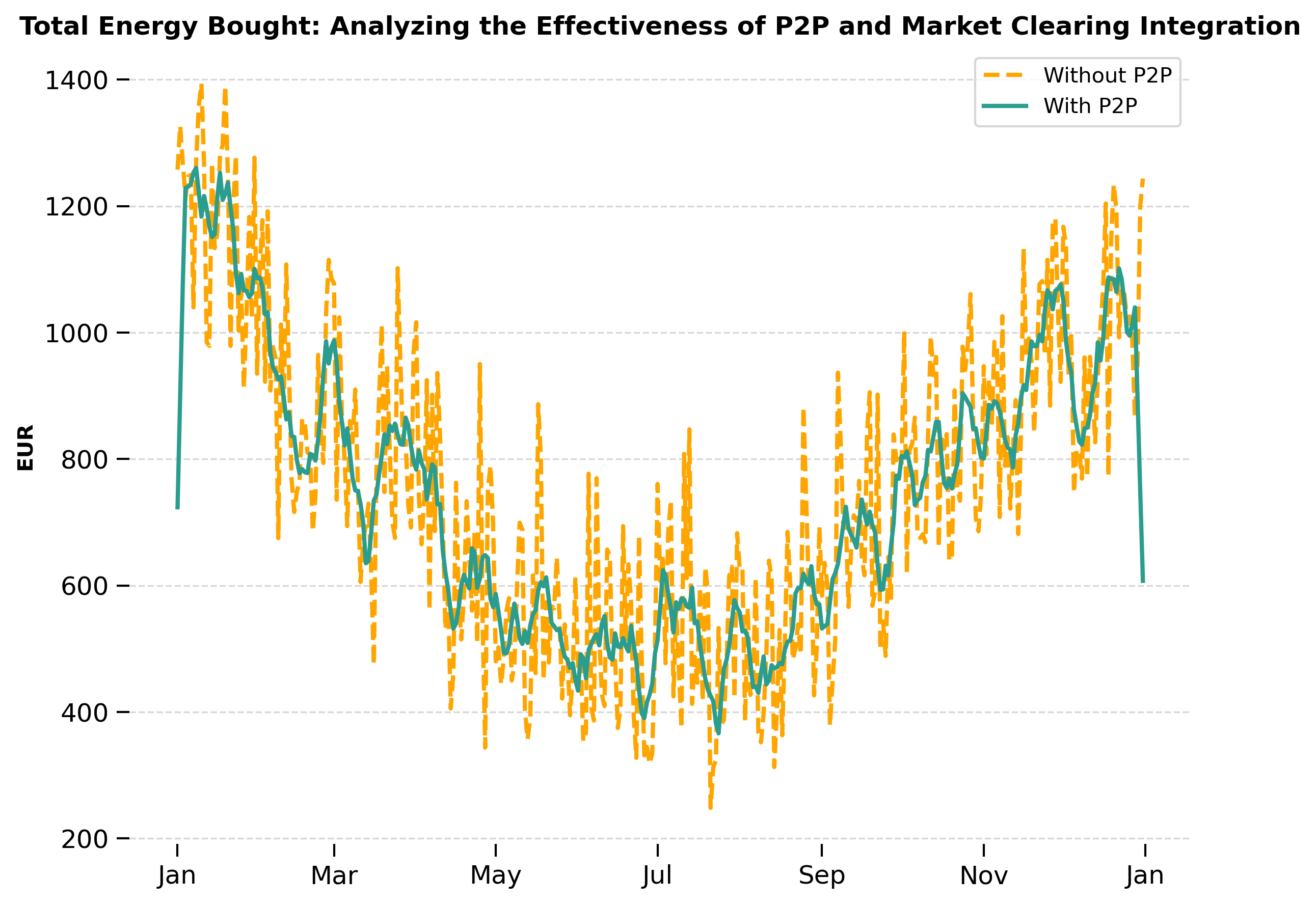}
        \caption{Comparison of Energy Purchased by Farms: P2P vs Non-P2P, all having Renewable Energy Sources}
        \label{boughtwithRE}
    \end{subfigure}

    \vspace{0.5cm}
    
    \begin{subfigure}{0.49\textwidth}
        \includegraphics[width=\linewidth]{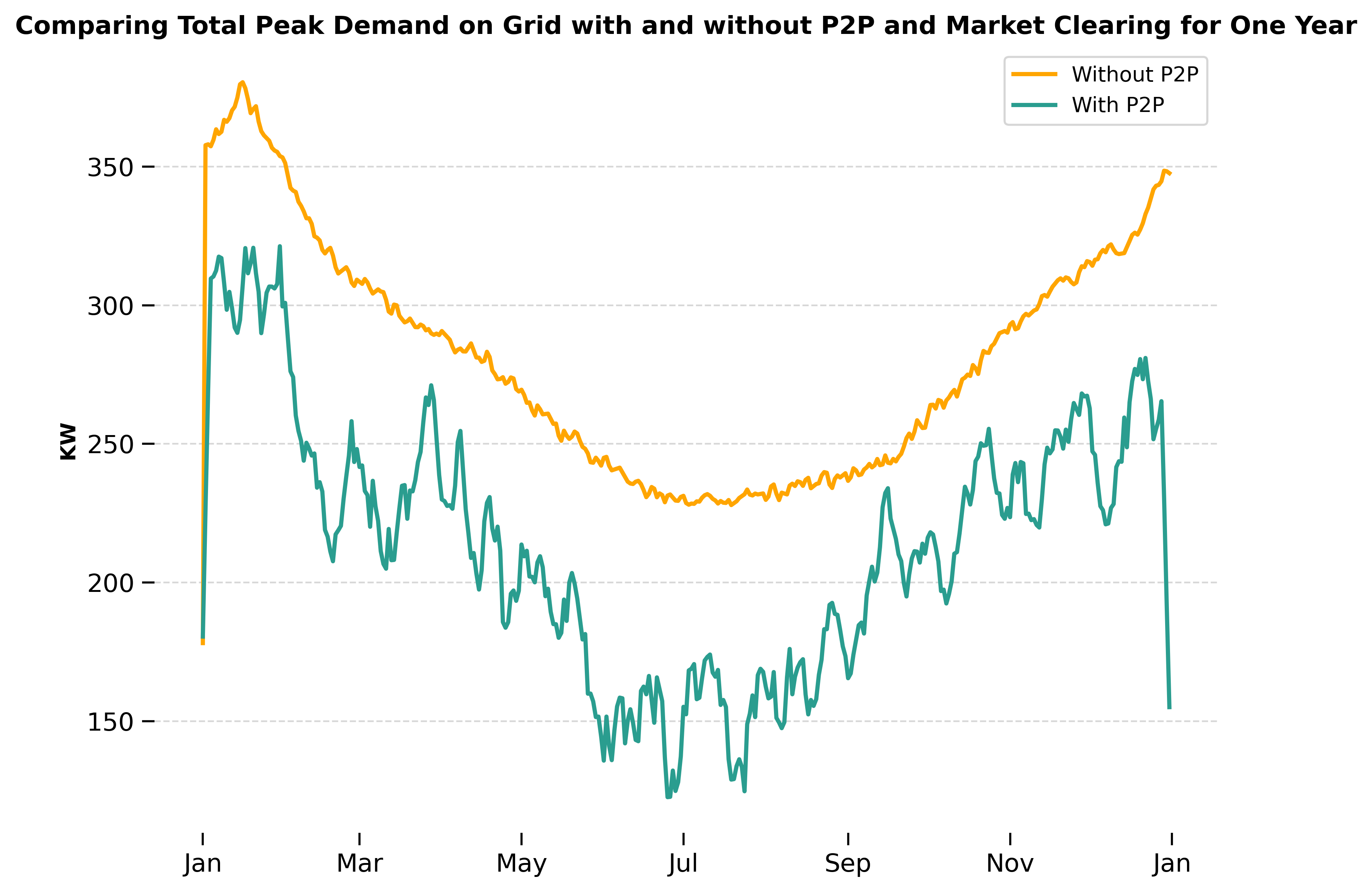}
        \caption{Comparison of Electricity Demand of Farms from Grid: P2P vs Non-P2P in Peak Hours}
        \label{peakdemand}
    \end{subfigure}
    \hfill
    \begin{subfigure}{0.49\textwidth}
        \includegraphics[width=\linewidth]{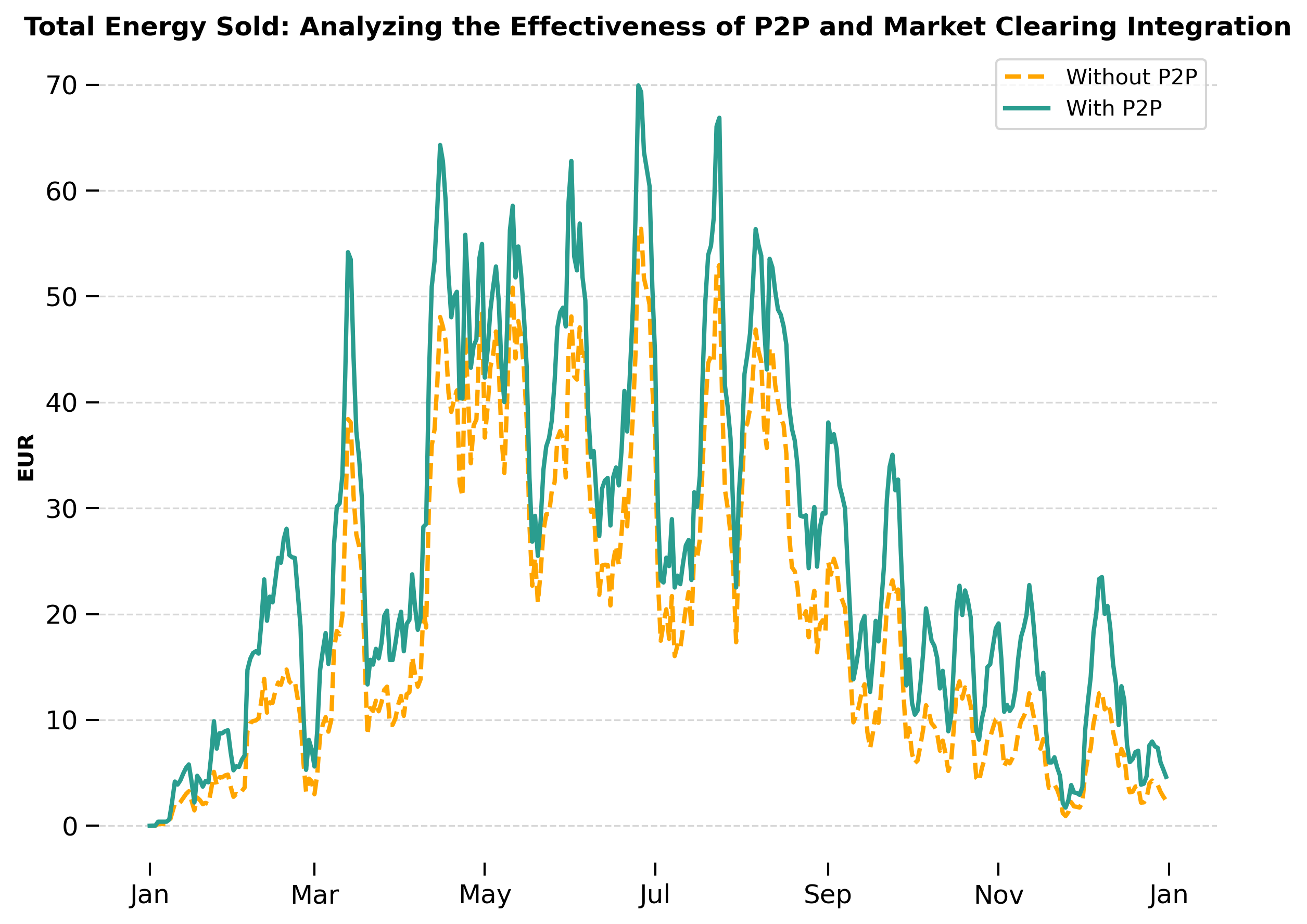}
        \caption{Comparison of Energy Sold by Farms: P2P vs Non-P2P}
        \label{EnergySold}
    \end{subfigure}
    
    \caption{Simulation Results}
    \label{fig:combined}
\end{figure}

\section{Experimental Results}
\label{results}

We evaluated our simulation approach by conducting a one-year simulation with a one-hour time step involving 10 prosumer farms. We assessed the results using five key metrics: (i) energy purchased by farms with and without RE sources and P2P trading, (ii) energy purchased by farms with P2P trading compared to no P2P trading, all with RE sources, (iii) energy sold by farms with RE sources (P2P vs. non-P2P), (iv) peak hour energy demand from the grid by farms with RE sources (P2P vs. non-P2P), and (v) effective ISP and IBP advice based on time-of-use (ToU) tariffs.

Figure \ref{boughtwoRE} compares the cost trend per day between farms without RE generation sources and P2P energy trading, and farms with RE generation sources and P2P energy trading. The results show that the community using RE sources and P2P energy trading paid approximately 30\% less compared to the farms without these features. Figure \ref{boughtwithRE} demonstrates the cost difference between farms with RE generation sources but no P2P trading, and farms with both RE sources and P2P trading. The results show a reduction of approximately 1\% in energy purchasing costs for the community. Figure \ref{peakdemand} illustrates the reduction in peak hour demand from the grid, with a 24\% decrease observed over the course of the year. Figure \ref{EnergySold} shows the potential earnings from selling excess energy, indicating that farms with P2P trading can earn approximately 37\% more compared to selling energy solely to the grid. These results highlight the cost savings and reduced reliance on the grid achieved through P2P energy trading, RE generation sources, and market clearing strategies.

\section{Conclusion and Future Work}

This research demonstrates the efficacy of integrating distributed P2P energy trading, RE resources, and auction-based market clearing mechanisms within a dairy farming community. The one-year simulation with 10 farms shows that P2P trading significantly reduces energy costs and reliance on the grid. The research contributions resulted in:
\\1) Farm community reduced energy purchases from the grid by 30\% with the use of RE sources and P2P trading compared to no RE sources and P2P trading.\\
2) With RE sources, the farm community increased profit from selling excess energy to peers and the grid by 37\% with P2P energy trading compared to no P2P trading.\\
3) With RE sources, farms' electricity demand from the grid, especially during peak hours, decreased by 24\% with P2P energy trading compared to no P2P trading.\\
4) Regular updates on market conditions were provided to participants, ensuring transparency and informed decision-making during auctions.\\
To improve the model, further research should investigate the influence of bidding prices on load and battery management and line losses. One potential approach is to employ multi-agent reinforcement learning (MARL) techniques to enhance market estimation and decision-making. These advancements can lead to more accurate models of market behavior and enable informed energy trading in dairy farms.

\section*{Acknowledgements}
This publication has emanated from research conducted with the financial support of Science Foundation Ireland under Grant number [21/FFP-A/9040].

%
%
%
%
\bibliography{paper}

\end{document}